\documentclass{article}

\usepackage{arxiv}

\usepackage[utf8]{inputenc} % allow utf-8 input
\usepackage[T1]{fontenc}    % use 8-bit T1 fonts
\usepackage{hyperref}       % hyperlinks
\usepackage{url}            % simple URL typesetting
\usepackage{booktabs}       % professional-quality tables
\usepackage{amsfonts}       % blackboard math symbols
\usepackage{nicefrac}       % compact symbols for 1/2, etc.
\usepackage{microtype}      % microtypography
\usepackage{lipsum}
\usepackage{graphicx}
\graphicspath{ {./images/} }

\usepackage{listings}
\usepackage{color}
\definecolor{dkgreen}{rgb}{0,0.6,0}
\definecolor{gray}{rgb}{0.5,0.5,0.5}
\definecolor{mauve}{rgb}{0.58,0,0.82}
\title{SBMLDiagrams: A python package to process and visualize SBML layout and render}

\author{
 Jin Xu \\
  Department of Bioengineering\\
  University of Washington\\
  Seattle, WA 98195 \\
   \And
 Jessie Jiang \\
  Department of Computer Science and Engineering\\
  University of Washington\\
  Seattle, WA 98195 \\
  \And
 Herbert M. Sauro \\
  Department of Bioengineering\\
  University of Washington\\
  Seattle, WA 98195 \\
}

\begin{document}
\maketitle
\begin{abstract}
\textbf{Summary:} The Systems Biology Markup Language (SBML) is an extensible standard format for exchanging biochemical models. One of the extensions for SBML is the SBML Layout and Render package. This allows modelers to describe a biochemical model as a pathway diagram. However, up to now there has been little support to help users easily add and retrieve such information from SBML. In this application note, we describe a new Python package called SBMLDiagrams. This package allows a user to add layout and render information or retrieve it using a straightforward Python API. The package uses skia-python to support the rendering of the diagrams, allowing export to commons formats such as PNG or PDF. 
\\
\textbf{Availability:} SBMLDiagrams is publicly available and licensed under the liberal MIT open-source license. The package is available for all major platforms. The source code has been deposited at GitHub (\url{github.com/sys-bio/SBMLDiagrams}). Users can install the package using the standard pip installation mechanism: {\tt pip install SBMLDiagrams}.\\
\textbf{Contact:} \href{hsauro@uw.edu}{hsauro@uw.edu}\\
\textbf{Supplementary information:} Supplementary data are available at \textit{Bioinformatics} online.
\end{abstract}

% keywords can be removed

\section{Introduction}
The Systems biological Markup Language (SBML) \cite{hucka2003systems} is a markup language to describe biochemical models of biological systems. SBML is used primarily to allow the exchange of models between different software tools. To read and write SBML, users are recommended to use the software library libSBML \cite{bornstein2008libsbml}. libSBML allows users to read, write and create SBML documents from a wide range of languages such as  C, C++, Java, and Python. libSBML also supports extensions such as SBML layout and render \cite{deckard2006supporting} which allows users to describe biochemical models as pathway diagrams. The layout is used to describe the positions of elements on a canvas while the render extension is used to describe how elements are rendered in graphical form. libSBML, however, can be difficult for new users to learn requiring fairly detailed knowledge of the layout and render object model. See Supplementary.

In this application note, we describe a python package called SBMLDiagrams that allows users to easily read, write, create and view layout and render specifications for a biochemical model without requiring an understanding of the underlying object model. The package makes use of python-libSBML, simplesbml \cite{sauro2020simplesbml}, and skia-python to add and inspect the layout and render part of SBML files as well as render the diagram to a variety of formats such as PNG and PDF. SBMLDiagrams support all specific layout elements.

\section{Methods}
%\label{sec:headings}
SBML level 3 supports layout and render information. The Layout describes the positions and sizes of different graphical objects, including compartments, species, and reactions. The render describes the color and shape information. SBMLDiagrams can be used to read, write, create and/or visualize SBML files based on the layout and render information. If an SBML file contains no layout or render information, SBMLDiagrams can be used to add this information. 
\subsection{Working with SBML Layout and Render}
%\lipsum[6]
SBMLDiagrams can be used to read and edit the layout and render of an existing SBML model which can be subsequently exported to an updated SBML model. A variety of color formats (as well as opacity) are supported including decimal RGB, HEX string, and all the HTML names (\url{https://www.w3schools.com/colors/colors_names.asp}). The following is an example that illustrates some simple operations.
\begin{lstlisting}
import SBMLDiagrams

# load SBML from a file (or string)
df = SBMLDiagrams.load('mymodel.xml')
print(df.getNodeSize ("ATP"))
print(df.getReactionFillColor ("ENOLASE"))
df.setNodeTextPosition ("ATP", [30, 30])
df.setNodeFillColor ("ATP", "Seagreen", opacity=0.5)
df.setNodeTextFontColor ("ATP", "#FF6347", opacity=1.)
df.setReactionLineThickness ("ENOLASE", 3.)
newSbmlStr = df.export()
\end{lstlisting}
\subsection{Visualization of SBML networks}
\begin{figure}%[!tpb]%figure1
  \centering
  \includegraphics[width=70mm,scale=0.5]{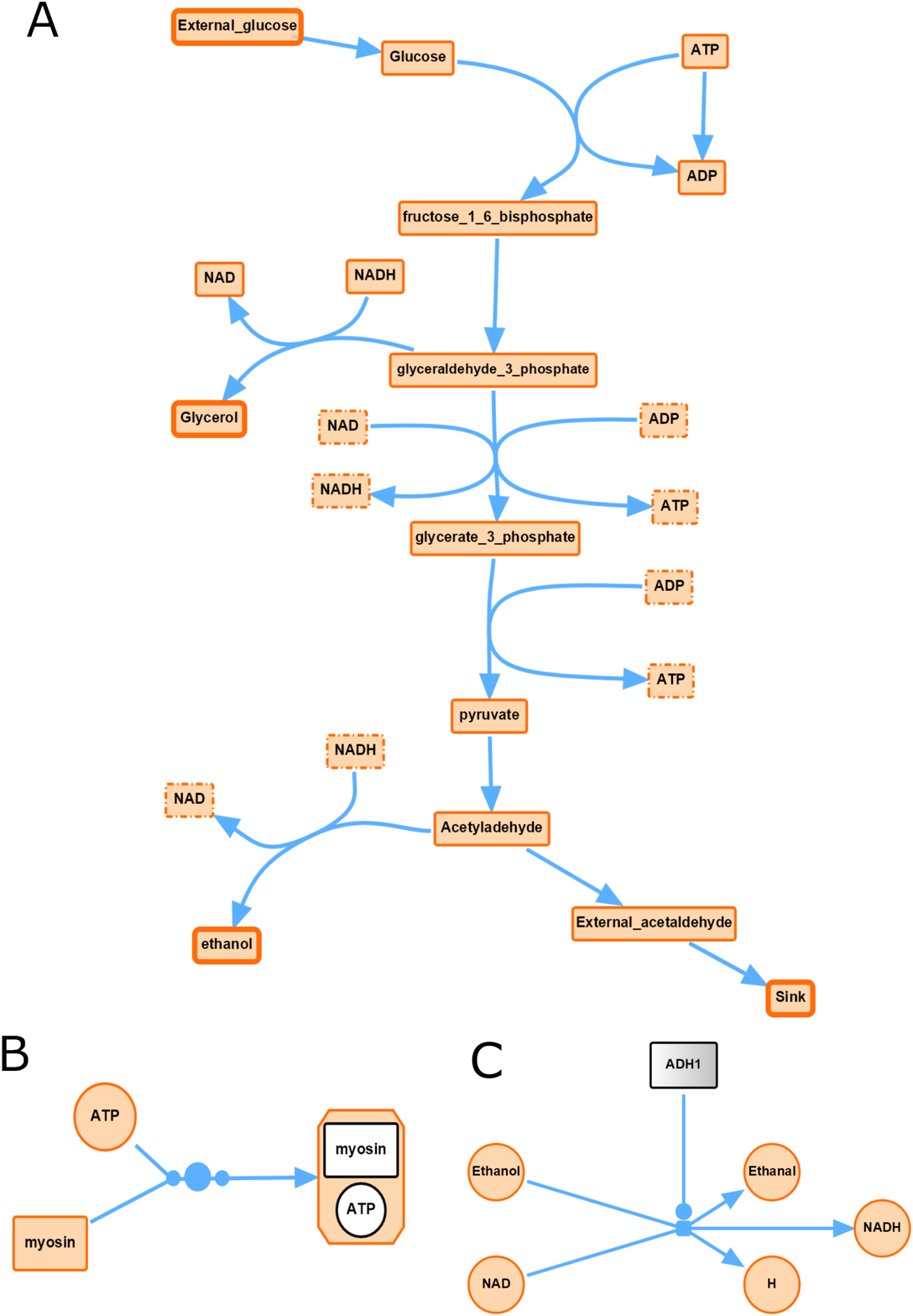}
  \caption{{\bf Some visualization examples by SBMLDiagrams.} {\bf A.} Using SBMLDiagrams to visualize a model of glycolysis \cite{wolf2001mathematical}. Alias nodes are indicated with dashed border lines and boundary nodes have a thicker border width than floating nodes. An animation is also available on Github (\url{https://github.com/sys-bio/SBMLDiagrams}). {\bf B.} Interface to SBGN with a complex species. {\bf C.} Interface to SBGN with a gradient node.}
  \label{fig:01}
\end{figure}

%\lipsum[6]
The visualization is based on skia-python The visualization is based on skia-python (\url{pypi.org/project/skia-python/}). Users can export the SBML networks as PNG, JPG or PDF files. Figure \ref{fig:01}A shows a visualization example of an SBML file from Jana Wolf's work \cite{wolf2001mathematical}. The following example illustrates the use of a modeling package such as Tellurium with SBMLDiagrams.\\
\begin{lstlisting}
import SBMLDiagrams, tellurium as te
r = te.loada ('''
A -> B; v; 
B -> C; v; 
C -> D; v;
v = 0
''')
df = SBMLDiagrams.load(r.getSBML())
df.autolayout()
df.draw()
\end{lstlisting}
For most Python IDEs {\tt df.draw()} will display the network at the console. Users can also save the figure to a file, for instance, {\tt df.draw(output\_fileName = 'fileName.pdf')}.

Autolayout is provided by NetworkX and is useful when an SBML model has no layout information. By use of drawing primitives it is also possible to specify networks that use some aspects from SBGN \cite{novere2009systems}, see Figure \ref{fig:01}B and \ref{fig:01}C. SBMLDiagrams can also be driven from Tellurium to create simulation animations overlaid onto a diagram.

%\subsection{Figures}
%\lipsum[10] 
%See Figure \ref{fig:fig1}. Here is how you add footnotes. \footnote{Sample %of the first footnote.}
%\lipsum[11] 

%\begin{figure}
  %\centering
  %\includegraphics{drawing350.eps}
  %\fbox{\rule[-.5cm]{4cm}{4cm} \rule[-.5cm]{4cm}{0cm}}
  %\caption{Sample figure caption.}
  %\label{fig:fig1}
%\end{figure}

\section{Implementation}
SBMLDiagrams is available as a pip package (\url{pypi.org/project/SBMLDiagrams/}), and installed using the standard pip mechanism: {\tt pip install SBMLDiagrams}. The package is fully documented as~\url{https://sys-bio.github.io/SBMLDiagrams/}.
\section*{Acknowledgements}
\textit{Funding}: This work was supported by the National Institutes of Health [U24EB028887].\\
\textit{Conflict of Interest}: none declared.\\
We thank Frank T. Bergmann for his assistance in using python-libSBML.\\
This article has been accepted for publication in Bioinformatics ©: [2022] [owner as specified on the article]. Published by Oxford University Press [\url{https://doi.org/10.1093/bioinformatics/btac730}]. All rights reserved.

%text text text\vspace*{-12pt}

%
\section*{Author Contributions}
J.X. was the main developer of the python package. J.J. did the interface to the animation, networkX and color styles. H.M.S was responsible for the conception, funding acquisition, project administration and software testing. J.X and H.M.S. wrote the manuscript. All authors reviewed the manuscript. 

\bibliographystyle{unsrt}  
%\bibliography{references.bib}


\begin{thebibliography}{00}
\bibitem{hucka2003systems}Hucka, M., Finney, A., Sauro, H., Bolouri, H., Doyle, J., Kitano, H., Arkin, A., Bornstein, B., Bray, D., Cornish-Bowden, A. \& Others The systems biology markup language (SBML): a medium for representation and exchange of biochemical network models. {\em Bioinformatics}. \textbf{19}, 524-531 (2003)
\bibitem{bornstein2008libsbml}Bornstein, B., Keating, S., Jouraku, A. \& Hucka, M. LibSBML: an API library for SBML. {\em Bioinformatics}. \textbf{24}, 880-881 (2008)
\bibitem{deckard2006supporting}Deckard, A., Bergmann, F. \& Sauro, H. Supporting the SBML layout extension. {\em Bioinformatics}. \textbf{22}, 2966-2967 (2006)
\bibitem{sauro2020simplesbml}Sauro, H. SimpleSBML: A Python package for creating, editing, and interrogating SBML models: Version 2.0. {\em ArXiv Preprint ArXiv:2009.01969}. (2020)
\bibitem{wolf2001mathematical}Wolf, J., Sohn, H., Heinrich, R. \& Kuriyama, H. Mathematical analysis of a mechanism for autonomous metabolic oscillations in continuous culture of Saccharomyces cerevisiae. {\em FEBS Letters}. \textbf{499}, 230-234 (2001)

\bibitem{novere2009systems}Novère, N., Hucka, M., Mi, H., Moodie, S., Schreiber, F., Sorokin, A., Demir, E., Wegner, K., Aladjem, M., Wimalaratne, S. \& Others The systems biology graphical notation. {\em Nature Biotechnology}. \textbf{27}, 735-741 (2009)
\end{thebibliography}

\end{document}